# Overview of Current Type I/II Kinase Inhibitors

Zheng Zhao[1], Philip E. Bourne[*,1,2]

*1*. Department of Biomedical Engineering, University of Virginia, Charlottesville, VA 22903, United States of America

*2*. Data Science Institute, University of Virginia, Charlottesville, VA 22904, United States of America

*Corresponding author

   Philip E. Bourne: phone, (434) 924-6867; e-mail, peb6a@virginia.edu

**Abstract.**

Research on kinase-targeting drugs has made great strides over the last 30 years and is attracting greater attention for the treatment of yet more kinase-related diseases. Currently, 42 kinase drugs have been approved by the FDA, most of which (39) are Type I/II inhibitors. Notwithstanding these advances, it is desirable to target additional kinases for drug development as more than 200 diseases, particularly cancers, are directly associated with aberrant kinase regulation and signaling. Here, we review the extant Type I/II drugs systematically to obtain insights into the binding pocket characteristics, the associated features of Type I/II drugs, and the mechanism of action to facilitate future kinase drug design and discovery. We conclude by summarizing the main successes and limitations of targeting kinase for the development of drugs.

**1. Introduction.**

Kinases are enzymes that phosphorylate specific substrates and in so doing play a vital role in signal transduction networks [1,25,27]. Clinical evidence has shown that aberrant kinase regulation and catalysis are directly associated with more than 200 diseases, especially cancer [6,23]. Thus, exploring the therapeutic potential of the human kinome is highly desirable for the treatment of many diseases [13,37,39,46]. As of August 2018, 42 kinase-targeted drugs have been approved by the U.S. Food and Drug Administration (FDA) [35]. Notwithstanding these advances, developing a desired kinase drug is still a challenging task due to the high similarity of the ATP binding sites across the whole kinome thwarting selectivity [21,29]. Here we systematically review extant kinase inhibitors, especially FDA-approved kinase drugs, to provide benchmarks and useful clues for prospective kinase drugs.

Over the last 30 years, chemically diverse kinase inhibitors with varied selectivity levels have emerged and can been classified into four types based on their binding modes: Type I, Type II, Type III and Type IV (Figure 1) [11,20,41,52]. Type I inhibitors, such as baricitinib (Figure 1a), occupy the ATP-bound pocket of the kinase in the "DFG-in" conformation. Type II kinase inhibitors, such as imatinib, not only occupy the ATP-bound pocket of the kinase with the "DFG-out" inactive conformation but elongate to the adjacent allosteric pocket (Figure 1b). Type III and Type IV inhibitors are also called allosteric inhibitors [40]. Type III inhibitors bind in the allosteric site close to the ATP-bound pocket, such as with the MEK inhibitor cobimetinib [32] (Figure 1c). Type IV inhibitors bind to the allosteric pocket distant from the ATP-binding site, such as the allosteric pocket at the C-lobe [45] (Figure 1d), the allosteric pockets on the surface of the kinase domain [16,42,47] (Figure 1e), or the allosteric pocket at the N-lobe [4] (Figure 1f). By the numbers, the vast majority of kinase inhibitors are Type I/II inhibitors. Of the 42 FDA-approved kinase drugs, 33 are Type I inhibitors, 6 are Type II inhibitors, and 3 MEK-targeted drugs are Type III inhibitors (Table 1). Here we focus on a review of the characteristics of current Type I/II kinase inhibitors.

## 2. FDA approved kinase inhibitors: Type I and II mechanisms of action.

### 2.1 Type I binding modes.

Type I inhibitors bind at the ATP-binding pocket, which is highly conserved across the human kinome [22,36,52]. To achieve greater selectivity than ATP, Type I inhibitors typically not only occupy the space where the ATP adenine group binds but also extends into different proximal regions, specifically referred to as the front-pocket region, the hydrophobic pocket region, the DFG motif, and the P-loop region (Figure 2a)[24,49]. Here we outline how inhibition occurs through the ability to combine the adenine-binding area with different proximal regions.

Gefitinib is one of the first-generation EGFR drugs for the treatment of non-small cell lung cancer (NSCLC) [33]. Gefitinib binds to EGFR by its quinazoline scaffold, which forms hydrogen bonds with the hinge region, mimicking the hydrogen bonds between the hinge region and the adenine moiety of ATP, one 4-position substitutional group extends into the hydrophobic pocket and one 6-position morpholine derivative extends into the front pocket and forms polar interactions with the adjacent residues C797 and D800 (Figure 2a) [12]. Similarly, the other 11 kinase drugs (Figure 2b), although acting on different kinase targets, also share the same binding pattern as gefitinib. This class of Type I binding, which consists of binding to the adenine-binding pocket, the hydrophobic pocket region, and the front pocket region, constitutes the largest cluster of 12 FDA approved drugs. In this class, it is worth noting that afatinib, neratinib, ibrutinib, and acalabrutinib not only contain the corresponding hydrophilic substituents elongating into the front pocket but also carry an acrylamide electrophilic group forming a covalent interaction with a nearby cysteine within the front pocket region [7,48,49]. The four covalent drugs together with another covalent drug osimertinib are irreversible kinase drugs, which are discussed in the chapter on irreversible inhibitors.

Ribociclib is an inhibitor of cyclin D1/CDK4 and CDK6 for the treatment of hormone receptor-positive, HER2-negative metastatic or advanced breast cancers [19]. Like the binding mode of gefitinib (Figure 2a), ribociclib binds to the adenine-binding area via its 2-amino pyrimidine scaffold forming two hydrogen bonds with the hinge and binds to the front-pocket region via the piperazine-substituent group (Figure 3a) [8]. However, ribociclib does not occupy the hydrophobic pocket like gefitinib but interacts with the DFG-motif region via one carboxamide group. The carboxamide group forms interactions with Asp163 of the DFG motif and Lys43, which is located on the "roof" of the ATP-binding site (Figure 3a). The other 9 drugs (Figure 3b) share the same binding mode as that of ribociclib (i.e., occupying the front-pocket region, the adenine-binding area and DFG-motif region), although targeting different kinases. In this class, encorafenib was approved by the FDA in combination with binimetinib for the treatment of patients in June 2018. The encorafenib/binimetinib combination has shown the best-in-class efficacy and tolerability for patients with BRAF V600E –mutant advanced unresectable or metastatic melanoma [10]. Similar drug combinations had been validated previously; dabrafenib/trametinib was approved in Jan. 2014 and vemurafenib/cobimetinib was approved in Nov. 2015 for the same population [9]. Using drug combinations represents an emerging kinase application strategy in clinical care, owing to a more detailed understanding of the underlying kinase signaling networks.

Vemurafenib exhibits a different binding mode. Vemurafenib works specifically for melanoma patients with the BRAF V600E mutation [28] and was approved by the FDA in August 2011. Vemurafenib binds to the kinase BRAF and contains a chlorobenzene group occupying the front-pocket region, the 7-azaindole group occupies the adenine-binding area and forms hydrogen-bond interactions, the sulfonamide group interacts with the DFG-motif region and the propyl group extends into the hydrophobic pocket region (Figure 4a) [5]. Due to this extension of the propyl group into the hydrophobic pocket, leading to the c-Helix-out displacement, BRAF adopts a DFG-in/c-Helix-out inactive conformation [34]. The drug lenvatinib follows the same binding mode as vemurafenib (Figure 4b) [30]. Unlike the other drugs described thus far, the binding modes of vemurafenib and lenvatinib avail themselves of additional adjacent regions to achieve selectivity, including the front-pocket region, the DFG-motif region and the hydrophobic pocket.

Midostaurin has been found to be active against more than 100 kinases and was approved for the treatment of adult patients with fms-like tyrosine kinase 3 (FLT3)-positive AML in combination with chemotherapy [38]. Midostaurin binds to the ATP-bound space of FLT3 (Figure

5a). Specifically, the pyrrolidine group of midostaurin forms two hydrogen bonds with the hinge region of FLT3 and the benzamidine group and the indole group interact with the P-loop cleft. Although midostaurin is approved, common side effects result from lack of specificity and hence binding to off-targets [15]. Four other drugs (Figure 5b), tofacitinib, idelalisib, baricitinib, and ruxolitinib, follow the same binding mode to bind to their corresponding kinase targets by also occupying the adenine-binding region and the P-loop region.

The drugs dabrafenib, osimertinib, and fostamatinib exhibit different binding modes, respectively (Figure 6a-c). Dabrafenib is an effective drug in the treatment of advanced melanoma patients with the BRAF V600E mutation [14]. Dabrafenib binds to the front pocket region, the adenine-binding area and the P-loop region (Figure 6a) [44]. However, dabrafenib resistance develops in the majority of patients after approximately six months of treatment. To overcome this resistance, the FDA approved the combination of dabrafenib and trametinib, a MEK inhibitor [51] for BRAF V600E/K-mutant metastatic/advanced melanoma, or as an adjunct treatment for BRAF V600E advanced patients following chemotherapy [26]. Osimertinib is the third-generation EGFR T790M inhibitor to treat metastatic/advanced NSCLC [2]. Selectivity is achieved by binding to the P-loop region, the adenine-binding region and the front pocket region of the EGFR binding pocket (Figure 6b). To improve the selectivity, the acrylamide group is incorporated into osimeritinib to form covalent interaction with the reside C797. In the clinic, drug resistance usually develops in about 10 months mainly due to the C797S mutation [31]. Fostamatinib is a Syk inhibitor for the treatment of chronic immune thrombocytopenia (ITP) and bears a 3,4,5-trimethoxyphenyl group at the front pocket region, a pyrimidine group occupying the adenine-binding area, a pyridine derivative occupying the P-loop region, and a phosphate group binds within the DFG-motif region (Figure 6c).

In summary, Type I drugs bind to the common adenine scaffold region and extend into adjacent regions. Figures 2-6 show that it is vital to utilize one or more adjacent regions to achieve the desired selectivity for specific kinases. In Type I binding mode another common feature has the DFG motif in an "In" conformational state, which means the side chain of the aspartic acid of the DFG motif points to the hinge region of kinases, often called the "active" kinase state.

**2.2 Type II binding modes.**

Kinase structure research has provided a wealth of information on conformational plasticity, a major factor to determine different binding modes [52]. The Type II binding mode was validated with the approval of the first drug, imatinib, in 2001. The co-crystal Abl-imatinib complex demonstrated that the Type II inhibitor bound not only to the ATP adenine group area but extended into the allosteric pocket with the benzamide substituent (Figure 7a). In contrast to the aforementioned Type I binding mode required for the inhibitors to be accommodated in the "DFG-in" conformational state, the Type II inhibitor induces a dramatic displacement of the "DFG" motif (Figure 7a) and consequently, the sidechain of the phenylalanine of the "DFG" motif flips and points to the hinge, referred to as the "DFG-out" conformation [11]. In addition, the "DFG-out" motif normally forms a hydrogen-bond interaction with the amino group of imatinib. To date, 6 Type II kinase drugs have been approved (Figure 7b). Their common binding modes, occupying the adenine-binding area, the DFG-out motif region, and the allosteric pocket region, have formed the typical Type II binding pattern. It is worth noting that Type II inhibitors, occupying the allosteric pocket region [17], are not intrinsically more selective than Type I inhibitors [50].

**3. Successes and limitations.**

Over the last 30 years, kinase drug research has made great progress, transforming kinase targets being from "undruggable" to highly tractable. Consequently, kinase-targeted drugs have revolutionized the treatments of human cancers such as NSCLC, melanoma, thyroid cancer, breast cancer, lymphoma, and leukemia, as well as rheumatoid arthritis, immune thrombocytopenia [13]. Moreover, additional promising kinase targets [39], for example, CDK7, CDK11, and DYRK1A, have been added to an expanding druggable kinome [11].

Beyond single drug-single target pharmacology lies progress in addressing multi-gene-driven diseases using multi-target kinase drugs. Hence, it is important to systematically verify the target spectrum of a given inhibitor across the whole kinome. Correspondingly, kinome profiling techniques, such as KinomScan and Kinativ [11], have been developed to test kinome-wide selectivity. As such, the kinome-centric view has led to a standard protocol as part of kinase drug R&D [39]. Moreover, kinase research benefits from our increased understanding of signaling networks and the pathology of human diseases, which provides support to increasing number of combinations studies in both preclinical and clinical settings [9]. Take drug development for NSCLC as an example. Drug resistance frequently occurs after treating NSCLC with EGFR inhibitors [18]. Besides mutations in the kinase domain, notably T790M, further drug-resistance arises through the dysregulation of the mesenchymal-epithelial transition factor (MET) [3]. Recently, focusing on the resistance mechanism, it was found that a drug cocktail strategy combining capmatinib (a MET inhibitor) with gefitinib (an EGFR inhibitor) is a promising treatment for patients with EGFR-mutated-MET-dysregulated, particularly MET-amplified, NSCLC [43].

Covalent drugs with reduced toxicity and favorable selectivity have rapidly emerged [48], as exemplified by the recent approval of five type I irreversible kinase inhibitors Kinase covalent drugs form covalent interactions with noncatalytic residues such as cysteine or lysine situated around the binding pocket [49]. This has revitalized interest in covalently targeting kinases, and other protein targets in general.

Notwithstanding, side effect of applying kinase inhibitor drugs due to off-targeting still induce serious toxic effects, even end-of-life [13]. As such, kinase inhibitor profiling needs to be augmented to a broader spectrum of possible targets and the consequences explored both in vitro and in vivo.

A further limitation is that current kinase research focuses on approximately 40 kinases, concentrated in a few targets, which include VEGFR (8 drugs approved), EGFR (6 drug approved), ABL (5 drugs approved) and ALK (4 drug approved) [35]. Yet more than 100 kinases are directly associated with over 200 diseases. Much has been done, but there is much left to do [11].

**Acknowledgements**

Thanks to Peng Wu for useful insights and corrections when reviewing this manuscript.

**Figure 1.** The binding modes of Type I-IV inhibitors. (a) Type I inhibitor bound to the ATP binding pocket (pdb 4w9x, the JAK-baricitinib complex); (b) Type II inhibitor occupying the ATP binding pocket and extending into the allosteric pocket (pdb 4bkj, the ABL-imatinib complex); (c) Type III inhibitor bound to the allosteric pocket close to the ATP-bound pocket (pdb 4an2, the MEK-cobimetinib complex); (d) Type IV allosteric inhibitor bound to the allosteric pocket of the C-lobe (pdb 3k5v, the Abl-GNF-2 complex); (e) Type IV allosteric inhibitor bound to the allosteric pocket at the interface of the AKT1 kinase and the PH domain (pdb 3o96, the AKT1-MK-2206 complex); (f) Type IV allosteric inhibitor bound to the allosteric pocket at the N-lobe (pdb 3py1, the CDK2-2AN complex).

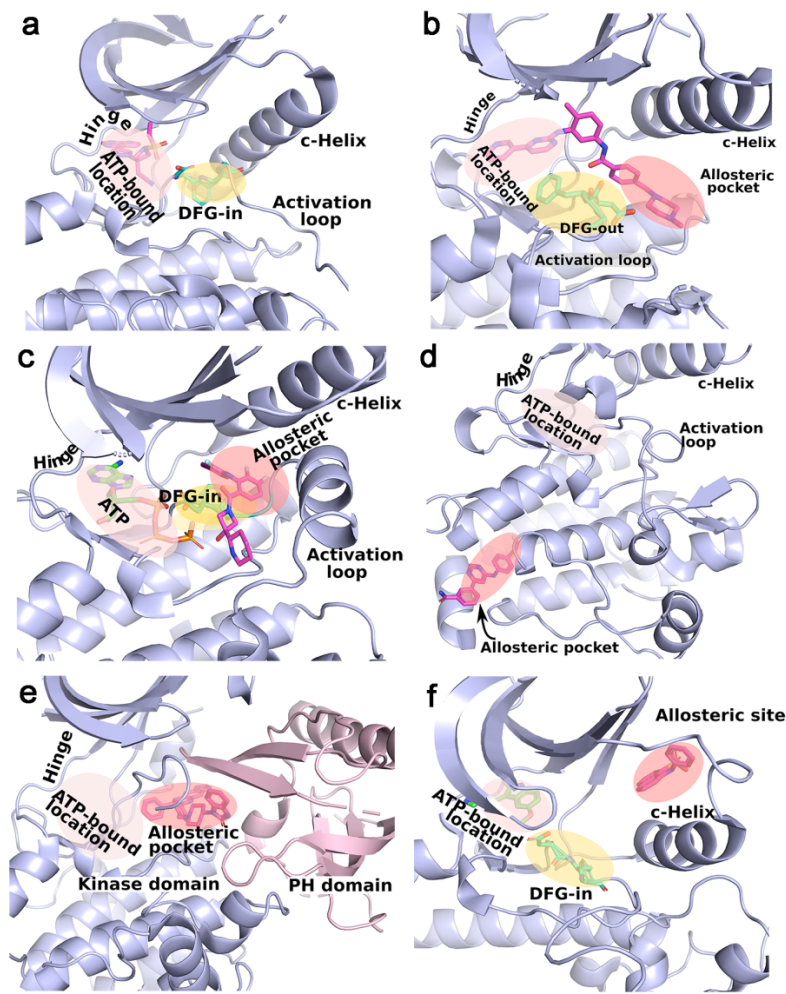

**Figure 2.** Type I inhibitors that occupy the front pocket region, the adenine-binding area, and the hydrophobic pocket region. (a) The drug gefitinib binding to the kinase EGFR (pdb 4I22). (b) The other 11 drugs with the same binding features as gefitinib.

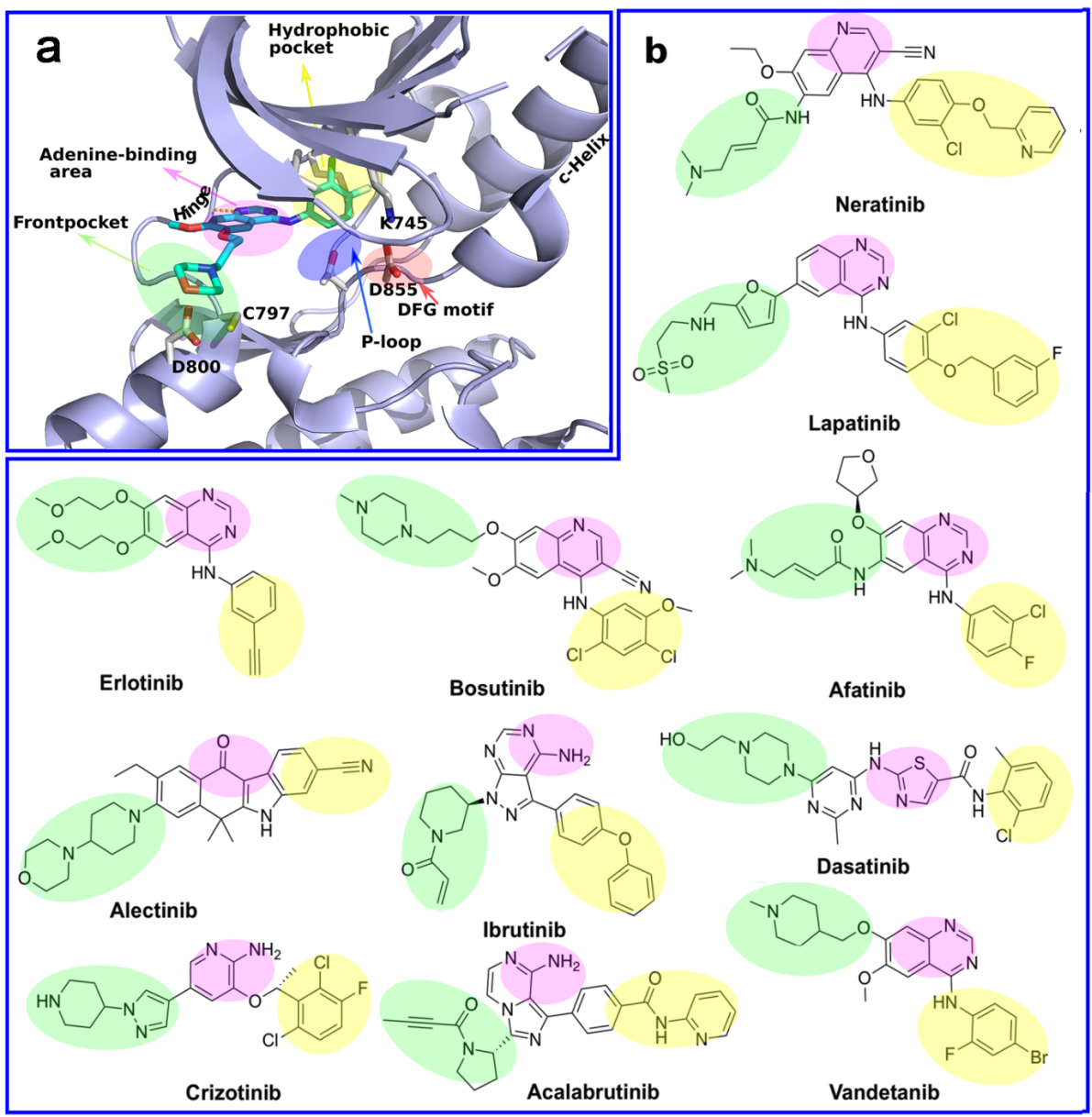

**Figure 3.** Type I inhibitors bound to the front pocket region, the adenine-binding area and the DFG-motif region. (a) The drug ribociclib binding to the kinase CDK6 (pdb 5l2t). (b) The other 9 drugs with the same binding features as ribociclib.

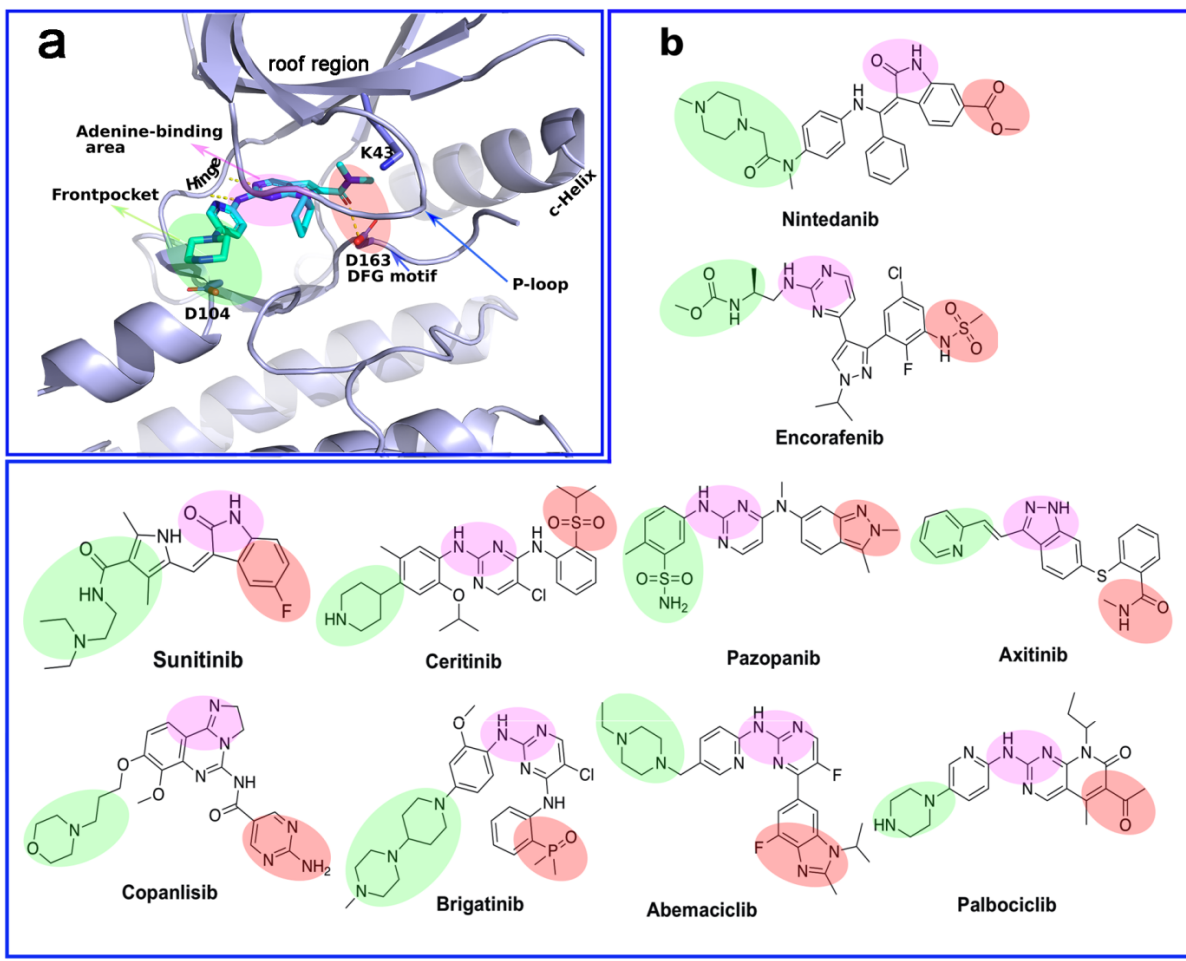

**Figure 4.** Type I inhibitors bound to the front pocket region, the adenine-binding area, the hydrophobic pocket region, and the DFG-motif region. (a) The kinase BRAF-vemurafenib complex (pdb 3og7). (b) The drug lenvatinib with the same binding features as vemurafenib.

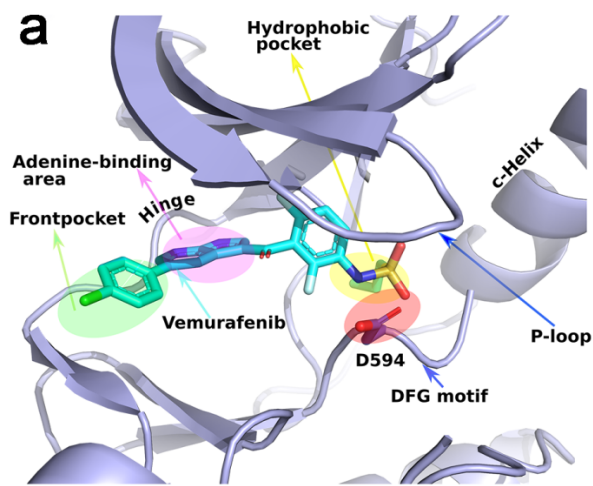
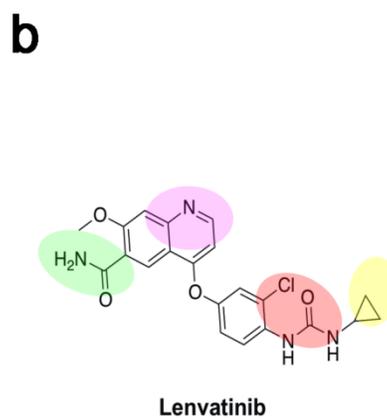

**Figure 5.** Type I inhibitors bound to the adenine-binding area and the P-loop region. (a) The kinase DYRK1A in complex with the drug midostaurin (pdb 4nct). (b) The other 4 drugs with the same binding features as midostaurin.

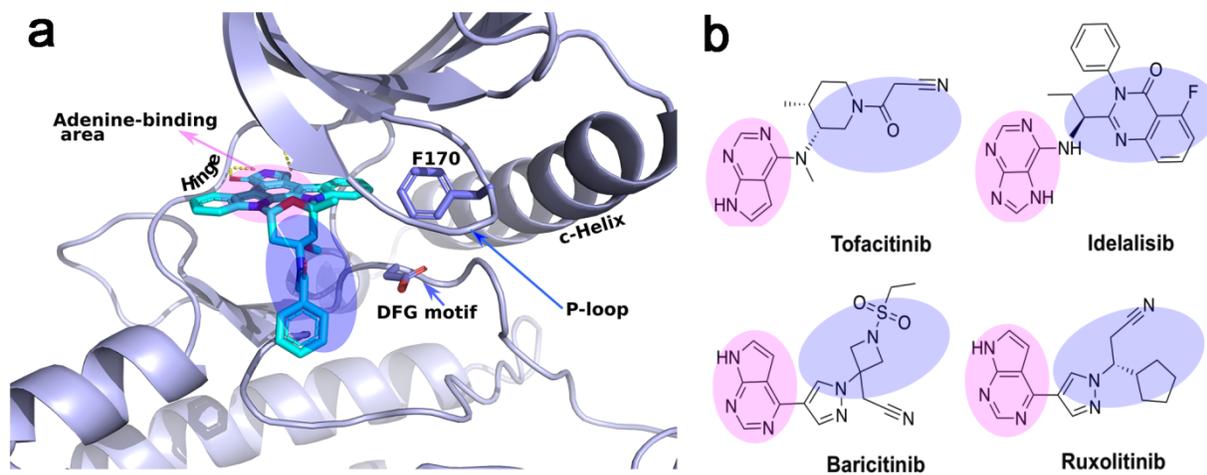

**Figure 6.** Type I kinase drug binding modes. (a) The drug dabrafenib bound to the adenine-binding area, the hydrophobic pocket region and the DFG-motif region of the kinase BRAF (pdb 4xv2). (b) The drug osimertinib bound to the front pocket region, the adenine-binding area and the P-loop region of the kinase EGFR (pdb 4zau). (c) The drug fostamatinib bound to the front pocket, the adenine-binding area, the DFG motif and the P-loop region of the kinase Syk (pdb 3fqs).

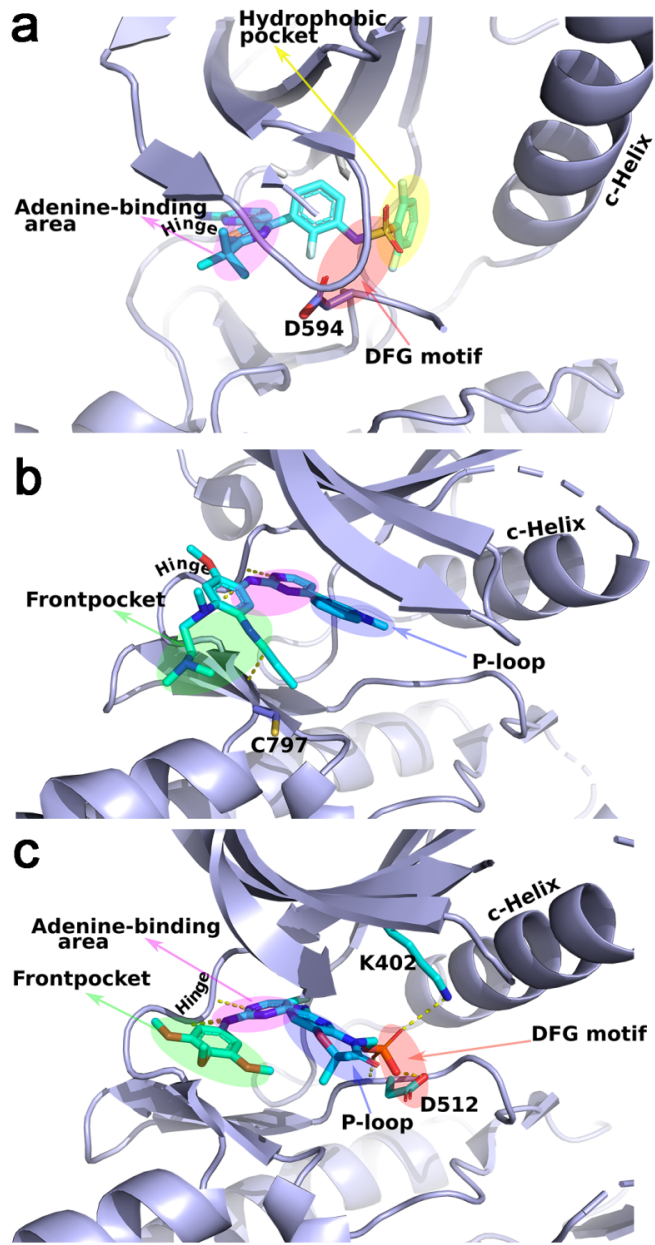

**Figure 7.** Type II inhibitors bound to the adenine-binding area, the DFG motif and the allosteric pocket region. (a) The Abl-imatinib co-crystal structure (pdb 1opj). (b) The other 5 drugs with the same binding features as imatinib.

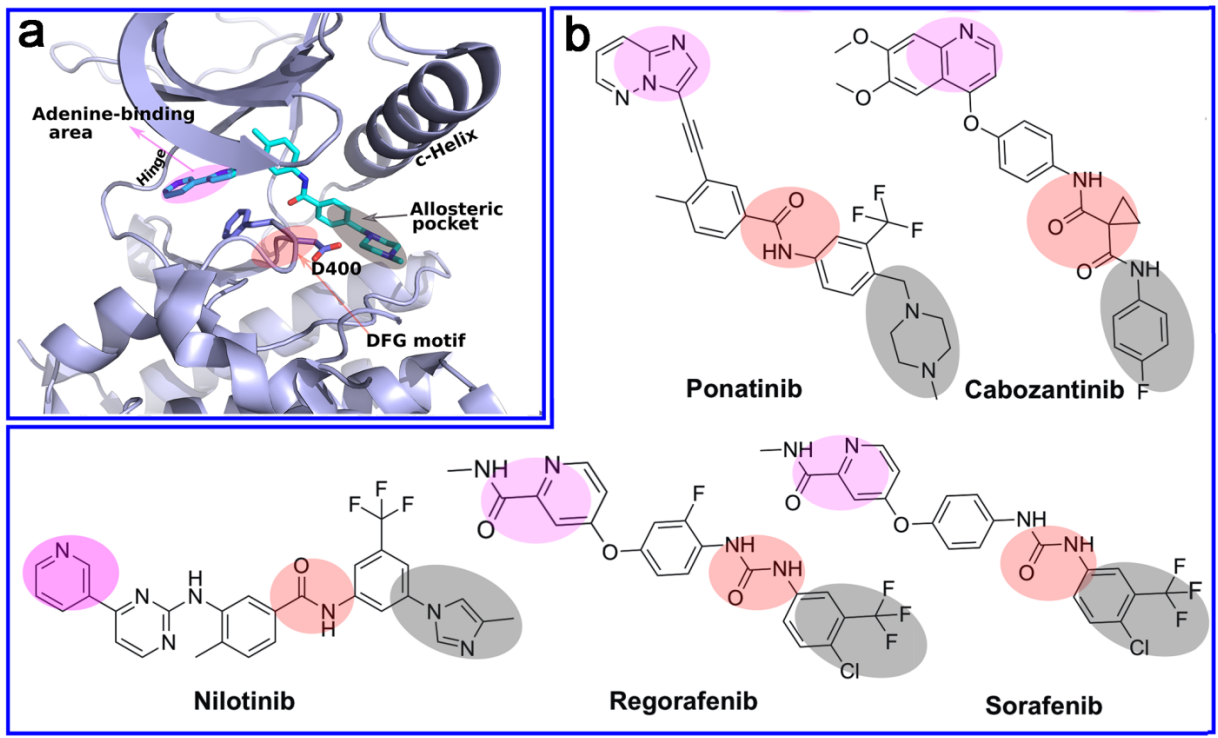

**Table 1**. FDA-approved kinase drugs with their associated binding modes and approval dates (as of August 2018).

| Approved Inhibitors | Approved Date | Primary Targets | Inhibitor of Type | PDB Entry | Reversible/ Irreversible |
|---|---|---|---|---|---|
| Imatinib | 2001/05 | ABL/PDGFR/c-KIT | II | 1OPJ | Reversible |
| Gefitinib | 2003/05 | EGFR | I | 4I22 | Reversible |
| Erlotinib | 2004/11 | EGFR | I | 4HJO | Reversible |
| Sorafenib | 2005/12 | VEGFR/PDGFR etc | II | 4ASD | Reversible |
| Sunitinib | 2006/01 | KIT/PDGFR etc | I | 2Y7J | Reversible |
| Dasatinib | 2006/06 | ABL/SRC etc | I | 3QLG | Reversible |
| Lapatinib | 2007/03 | EGFR/Her2 | I | 1XKK | Reversible |
| Nilotinib | 2007/10 | ABL/KIT etc | II | 3GP0 | Reversible |
| Pazopanib | 2009/10 | c-KIT/FGFR etc | I | - | Reversible |
| Vandetanib | 2011/04 | VEGFR/EGFR etc | I | 2IVU | Reversible |
| Crizotinib | 2011/08 | ALK/ROS1 | I | 3ZBF | Reversible |
| Vemurafenib | 2011/08 | BRAF | I | 3OG7 | Reversible |
| Ruxolitinib | 2011/11 | JAK1/JAK2 | I | 4U5J | Reversible |
| Axitinib | 2012/01 | VEGFR etc | I | 4AGC | Reversible |
| Bosutinib | 2012/09 | ABL/SRC | I | 4OTW | Reversible |
| Regorafenib | 2012/09 | VEGFR etc | II | - | Reversible |
| Tofacitinib | 2012/11 | JAK1/JAK3 | I | 3LXN | Reversible |
| Cabozantinib | 2012/11 | c-MET/VEGFR2 etc | II | - | Reversible |
| Ponatinib | 2012/12 | ABL | II | 4C8B | Reversible |
| Trametinib | 2013/05 | MEK1 | III | - | Reversible |
| Dabrafenib | 2013/05 | BRAF | I | 4XV2 | Reversible |
| Afatinib | 2013/07 | EGFR | I | 4G5J | Irreversible |
| Ibrutinib | 2013/11 | BTK | I | 4IFG | Irreversible |
| Ceritinib | 2014/04 | ALK | I | 4MKC | Reversible |
| Idelalisib | 2014/07 | PI3Kd | I | 4XE0 | Reversible |
| Nintedanib | 2014/10 | VEGFR etc | I | 3C7Q | Reversible |
| Palbociclib | 2015/02 | CDK4/CDK6 | I | 2EUF | Reversible |
| Lenvatinib | 2015/02 | VEGFR1/2/3 | I | 3WZD | Reversible |
| Cobimetinib | 2015/11 | MEK | III | 4AN2 | Reversible |
| Osimertinib | 2015/11 | EGFR | I | 4ZAU | Irreversible |
| Alectinib | 2015/12 | ALK | I | 5XV7 | Reversible |
| Ribociclib | 2017/03 | CDK4/CDK6 | I | 5L2T | Reversible |
| Brigatinib | 2017/04 | ALK/EGFR | I | 5J7H | Reversible |
| Midostaurin | 2017/04 | FLT3 etc | I | 4NCT | Reversible |
| Neratinib | 2017/06 | EGFR/HER2 | I | 2JIV | Irreversible |
| Abemaciclib | 2017/09 | CDK4/CDK6 | I | 5L2S | Reversible |
| Copanlisib | 2017/09 | PI3Ka/PI3Kd | I | 5G2N | Reversible |
| Acalabrutinib | 2017/10 | BTK | I | - | Irreversible |
| Fostamatinib | 2018/04 | SYK | I | 3FQS | Reversible |
| Baricitinib | 2018/05 | JAK1/2 | I | 4W9X | Reversible |
| Binimetinib | 2018/06 | MEK | III | 4U7Z | Reversible |
| Encorafenib | 2018/06 | BRAF | I | - | Reversible |